\documentclass[11pt]{amsart}

\usepackage[hmargin={3cm,3cm},vmargin={3cm,3cm},includehead]{geometry}
\usepackage{cite}
\usepackage{amsmath,amssymb}
\usepackage{amsfonts}
\usepackage[mathscr]{eucal}
\usepackage{dsfont}

\usepackage{epic}
\usepackage{eepic}

\newcommand{\ds}{\displaystyle}
\def\EXP{\textrm{{\large e}}}
\newcommand{\x}{\boldsymbol{x}}
\newcommand{\e}{\boldsymbol{e}}
\newcommand{\nop}{\boldsymbol{n}}
\newcommand{\ii}{\mathsf{i}}
\newcommand{\uop}{\boldsymbol{u}}
\newcommand{\wop}{\boldsymbol{w}}
\newcommand{\xop}{\boldsymbol{x}}
\newcommand{\pop}{\boldsymbol{p}}

\begin{document}

\vspace{2cm}

\title[]{Geometry of quadrilateral nets:\\ second Hamiltonian form.}%

\author[S. Sergeev]{Sergey M. Sergeev}

\address{Faculty of Informational Sciences and Engineering,
University of Canberra, Bruce ACT 2601}

\email{sergey.sergeev@canberra.edu.au}



\begin{abstract}
Discrete Darboux-Manakov-Zakharov systems possess two distinct
Hamiltonian forms. In the framework of discrete-differential
geometry one Hamiltonian form appears in a geometry of circular
net. In this paper a geometry of second form is identified.
\end{abstract}

\maketitle

The circular net \cite{KS98} --  a special type of
three-dimensional quadrilateral net \cite{DoliwaSantini} -- is an
example of \emph{geometrically integrable} (see \cite{BoSurBook}
and references therein) system endowed by a discrete space-time
Hamiltonian structure \cite{circular} what brings together
geometrically integrable and \emph{completely integrable}
Hamiltonian systems. A class of analytical equations describing
the three-dimensional quadrilateral nets is usually refereed to as
discrete Darboux-Manakov-Zakharov systems
\cite{Darboux,ZakharovManakov,BogdanovKonopelchenko}. In this
paper we discuss another special type of quadrilateral net whose
geometry is described by the second Hamiltonian form of DMZ
systems \cite{Melbourne}.

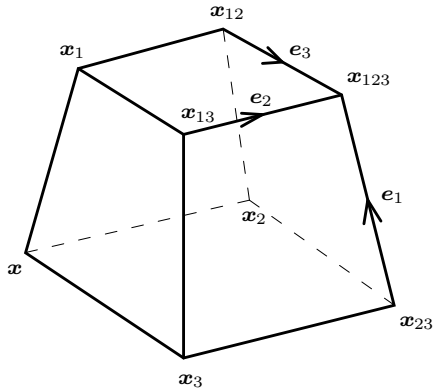
\begin{figure}[ht]
\begin{center}
\setlength{\unitlength}{0.35mm}
\begin{picture}(170,150)
\put(10,10){\begin{picture}(150,125)
\Thicklines
 \path(0,40)(60,0)(140,20)(120,100)(75,125)(20,110)(0,40)
 \path(60,85)(20,110)
 \path(60,85)(120,100)
 \path(60,85)(60,0)
 \path(128.5,51)(130,60)(135,52)
 \path(92,118)(97.5,112.5)(90,112)
 \path(82,93)(90,92.5)(82,88)
 \put(135,60){\scriptsize $\e_1$}
 \put(100,115){\scriptsize $\e_3$}
 \put(85,97){\scriptsize $\e_2$}
\thinlines
 \dashline{5}(85,60)(0,40)
 \dashline{5}(85,60)(75,125)
 \dashline{5}(85,60)(140,20)
 \put(59,91){\scriptsize $\x_{13}$}
 \put(58,-10){\scriptsize $\x_3$}
 \put(-7,32){\scriptsize $\x$}
 \put(142,12){\scriptsize $\x_{23}$}
 \put(13,115){\scriptsize $\x_1$}
 \put(122,105){\scriptsize $\x_{123}$}
 \put(70,130){\scriptsize $\x_{12}$}
 \put(82,52){\scriptsize $\x_2$}
\end{picture}}
\end{picture}
\end{center}
\caption{A cell of quadrilateral net.}\label{fig-cube}
\end{figure}

Following \cite{DoliwaSantini}, the $3D$ quadrilateral net is a
$\mathbb{Z}^3$ lattice imbedded into a multidimensional linear
space,
\begin{equation}
(n_1,n_2,n_3)\in\mathbb{Z}^3\;\to
\;\x(n_1,n_2,n_3)\in\mathbb{R}^M\;,\quad M\geq 3\;,
\end{equation}
such that each quadrilateral, e.g. {\small
\begin{equation}
\x=\x(n_1,n_2,n_3),\; \x_1=\x(n_1+1,n_2,n_3),\;
\x_2=\x(n_1,n_2+1,n_3),\; \x_{12}=\x(n_1+1,n_2+1,n_3),
\end{equation}}
is the planar one. A local cell (hexahedron) of quadrilateral net
is shown in Fig. \ref{fig-cube}.
Geometric integrability is based on the axiomatic statement
\cite{DoliwaSantini}: given the points
$\x_1,\x_2,\x_3,\x_{12},\x_{13},\x_{23}$ of the hexahedron, its
corners $\x$ and $\x_{123}$ can be obtained uniquely by a
two-dimensional ruler (ruler which draws a plane via three
non-collinear points).

The circular net is the quadrilateral net such that each its
hexahedron can be inscribed into a sphere. In this paper, instead
of the circular condition, suppose firstly that the target space
is four-dimensional Euclidean space,
\begin{equation}
\textrm{Q. net:}\;\;\mathbb{Z}^3\;\to\;\mathbb{E}_4\;.
\end{equation}
Each hexahedron is an element of a three-dimensional hyperplane.
In the framework of discrete-differential geometry
\cite{BoSurBook}, the quadrilateral net can be viewed as a planar
mesh of three-dimensional manifold embedded into four-dimensional
space. The hyperplanes are more general objects then
quadrilaterals since such net is not necessarily quadrilateral.

\begin{figure}[ht]
\setlength{\unitlength}{0.35mm} \centering
\begin{picture}(350,140)
\put(20,20){\begin{picture}(100,100)
 \Thicklines
 \path(50,50)(60,100)\path(52,70)(55,75)(56,69)
 \put(60,105){\scriptsize $\e_1'$}
 \path(50,50)(100,40)\path(69,43)(75,45)(70,48)
 \put(105,37){\scriptsize $\e_2'$}
 \path(50,50)(10,70)\path(23,66)(30,60)(20,62)
 \put(0,73){\scriptsize $\e_3$}
 \path(50,50)(50,0)\path(48,20)(50,25)(52,20)
 \put(50,-10){\scriptsize $\e_1$}
 \path(50,50)(0,40)\path(20,42)(25,45)(19,46)
 \put(-10,37){\scriptsize $\e_2$}
 \path(50,50)(90,20)\path(64,36)(70,35)(65,42)
 \put(93,15){\scriptsize $\e_3'$}
\end{picture}}
\put(200,10){\begin{picture}(150,125)
\Thicklines
 \path(0,40)(60,0)(140,20)(120,100)(75,125)(20,110)(0,40)
 \path(60,85)(20,110)
 \path(60,85)(120,100)
 \path(60,85)(60,0)
\thinlines
 \dashline{5}(85,60)(0,40)
 \dashline{5}(85,60)(75,125)
 \dashline{5}(85,60)(140,20)
 \put(59,91){\scriptsize $f$}
 \put(58,-10){\scriptsize $d$}
 \put(-7,32){\scriptsize $h$}
 \put(142,12){\scriptsize $e$}
 \put(13,115){\scriptsize $b$}
 \put(122,105){\scriptsize $a$}
 \put(70,130){\scriptsize $g$}
 \put(82,52){\scriptsize $c$}
\end{picture}}
\end{picture}
\caption{On the left: node $\x_{123}$ from Fig. \ref{fig-cube},
the junction of eight hyperplanes. On the right: dual graph to the
left vertex, corners $a,...,h$ label the hyperplanes.}
\label{fig-vertex}
\end{figure}
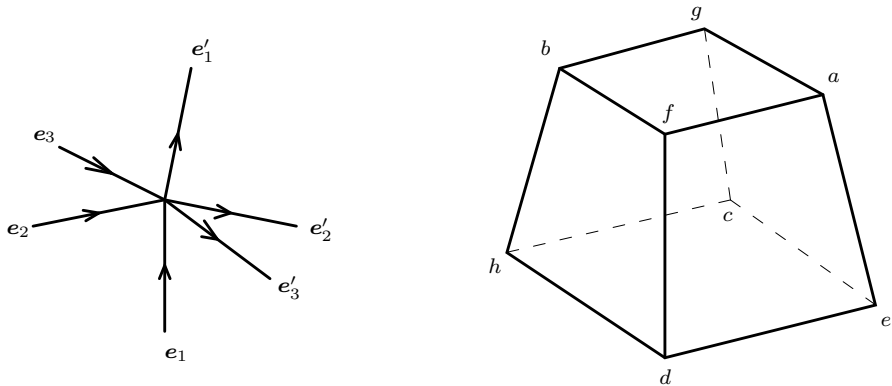

Let $\e_1,\e_2,\e_3$,
\begin{equation}
\e_1\;=\;\frac{\x_{123}-\x_{23}}{|\x_{123}-\x_{23}|}\;,\quad
\textrm{etc.,}
\end{equation}
be unit vectors defining the orientation of hexahedron in Fig.
\ref{fig-cube}, and let
\begin{equation}
\nop\;=\;\frac{*(\e_1\wedge\e_2\wedge\e_3)}{V(\e_1,\e_2,\e_3)}\;,\quad
\textrm{in indices:}\quad  (\nop)_\alpha =
\frac{\epsilon_{\alpha\beta\gamma\delta}(\e_1)^\beta (\e_2)^\gamma
(\e_3)^\delta}{V(\e_1,\e_2,\e_3)}\;,
\end{equation}
be the unit normal vector to the hyperplane $(\e_1,\e_2,\e_3)$.
Here
\begin{equation}
V(\e_1,\e_2,\e_3)\;=\;\textrm{volume of parallelipiped with the
edges $(\e_1,\e_2,\e_3)$}\;.
\end{equation}
Consider now node $\x_{123}$ of the net: the junction of eight
hyperplanes shown on the left  of Fig. \ref{fig-vertex}.
This junction is the subject of an extra ``orthogonality''
condition:
\begin{equation}\label{orthogonality}
\frac{V(\e_1,\e_2,\e_3)V(\e_1,\e_2',\e_3')V(\e_1',\e_2,\e_3')V(\e_1',\e_2',\e_3)}
{V(\e_1',\e_2',\e_3')V(\e_1',\e_2,\e_3)V(\e_1,\e_2',\e_3)V(\e_1,\e_2,\e_3')}\;=\;1\;.
\end{equation}
In some sense this condition is analogues to extra condition for
the circular net.

It is convenient to label the ``octants'' on the left of Fig.
\ref{fig-vertex} by corners of dual cube, see right part of Fig.
\ref{fig-vertex}. For instance,
\begin{equation}
\nop_h\sim *(\e_1\wedge \e_2\wedge \e_3)\;,\quad \nop_d\sim
*(\e_1\wedge \e_2\wedge \e_3')\;,\quad \nop_e\sim *(\e_1\wedge
\e_2'\wedge\e_3')\;,\quad \textrm{etc.}
\end{equation}
Orientation of each hyperplane is $\e_1^{\#}\wedge \e_2^{\#}
\wedge \e_3^{\#}$.

Consider now four hyperplanes $\nop_c$, $\nop_e$, $\nop_h$ and
$\nop_d$ surrounding the edge $\e_1$. Evidently, four hyperplanes
in four-dimensional linear space have a common edge if their
normal vectors are linearly dependent,
\begin{equation}\label{lp}
\nop_c - u_1\cdot\nop_e - w_1\cdot\nop_h +
\varkappa_1u_1w_1\cdot\nop_d\;=\;0\;.
\end{equation}
Numerical coefficients $u_1,w_1,\varkappa_1$ in (\ref{lp}) are
associated with the edge $\e_1$ which is orthogonal to all
$\nop_c,\nop_e,\nop_h$ and $\nop_d$. Analogous relations for edges
$\e_2$ and $\e_3$ are respectively
\begin{equation}\label{lp2}
\begin{array}{l}\ds \nop_h - u_2\cdot \nop_d - w_2\cdot \nop_b +
\varkappa_2u_2w_2\cdot \nop_f=0\;,\\ \ds \nop_c - u_3\cdot \nop_h
- w_3\cdot \nop_g + \varkappa_3u_3w_3\cdot \nop_b=0\;,\end{array}
\end{equation}
and such equations for outgoing edges $\e_i'$ are
\begin{equation}\label{lp3}
\begin{array}{l}\ds \nop_g - u_1'\cdot\nop_a - w_1'\cdot\nop_b +
\varkappa_1'u_1'w_1'\cdot\nop_f = 0\;,\\
\ds \nop_c - u_2'\cdot\nop_e - w_2'\cdot\nop_g + \varkappa_2'u_2'w_2'\cdot\nop_a=0\;,\\
\ds \nop_e - u_3'\cdot\nop_d - w_3'\cdot\nop_a +
\varkappa_3'u_3'w_3'\cdot \nop_f=0\;.\end{array}
\end{equation}
All numerical coefficients $u_i^{\#}$, $w_i^{\#}$ and
$\varkappa_i^{\#}$ can be expressed in terms of angular data as
follows. Let $\theta_{ce}$ be an angle between $\nop_c$ and
$\nop_e$,
\begin{equation}
(\nop_c,\nop_e)=\cos\theta_{ce}\;.
\end{equation}
Let further $\varphi_{1,e}$ be a dihedral angle of hyperplane
$\nop_e$ for the edge $\e_1$. In terms of unit vectors of Fig.
\ref{fig-vertex}, $\varphi_{1,e}$ is the dihedral angle between
planes $(\e_1,\e_2')$ and $(\e_1,\e_3')$. We extend
straightforwardly these self-explanatory notations to whole dual
graph of the junction, Fig. \ref{fig-vertex}. Then the
coefficients in relation (\ref{lp}) are given by
\begin{equation}\label{geometry}
u_1=\frac{\sin\varphi_{1,h}}{\sin\varphi_{1,d}}\frac{\sin\theta_{ch}}{\sin\theta_{ed}}\;,\quad
w_1=\frac{\sin\varphi_{1,e}}{\sin\varphi_{1,d}}\frac{\sin\theta_{ce}}{\sin\theta_{dh}}\;,\quad
\varkappa_1=\frac{\sin\varphi_{1,c}\sin\varphi_{1,d}}{\sin\varphi_{1,e}\sin\varphi_{1,h}}\;,
\end{equation}
and similarly for all other relations and their coefficients. The
geometry of junction without condition (\ref{orthogonality})
provides
\begin{equation}\label{kappas}
\varkappa_1'\varkappa_2'=\varkappa_1^{}\varkappa_2^{}\;,\quad
\varkappa_2'\varkappa_3'=\varkappa_2^{}\varkappa_3^{}\;.
\end{equation}
Since there are at most four linearly independent vectors among
eight $\nop_a,...,\nop_h$, the consistency of equations
(\ref{lp}-\ref{lp3}) relates the fields $u_i',w_i'$ on outgoing
edges and fields $u_i,w_i$ on incoming edges of Fig.
\ref{fig-vertex} as follows (see e.g. \cite{Sergeev:1999jpa}):
\begin{equation}\label{A-map}
\begin{array}{lll}
\ds u_1'=\Lambda_2^{-1}w_3^{-1}\;, & \ds
u_2'=\Lambda_1^{-1}u_3^{}\;, & \ds u_3'=\Lambda_1^{}u_2^{}\;,\\
[1mm] \ds w_1'=\Lambda_3^{}w_2\;, & \ds
w_2'=\Lambda_3^{-1}w_1^{}\;, & w_3'=\Lambda_2^{-1}u_1^{-1}\;,
\end{array}
\end{equation}
where
\begin{equation}\label{Lambda}
\begin{array}{l}
\ds
\Lambda_1^{}=u_1^{-1}u_3^{}-u_1^{-1}w_1^{}+\varkappa_1^{}w_1^{}u_2^{-1}\;,\\
[1mm] \ds \Lambda_2^{}=\frac{\varkappa_1}{\varkappa_2'}
u_2^{-1}w_3^{-1} + \frac{\varkappa_3}{\varkappa_2'}
u_1^{-1}w_2^{-1} - \frac{\varkappa_1\varkappa_3}{\varkappa_2'}
u_2^{-1}w_2^{-1}\;,\\
[1mm] \ds \Lambda_3^{}=w_1^{}w_3^{-1} -  u_3^{}w_3^{-1} +
\varkappa_3 w_2^{-1}u_3^{}\;.
\end{array}
\end{equation}
The ``orthogonality'' condition (\ref{orthogonality}) provides
\begin{equation}\label{kappa-inv}
\varkappa_i^{}=\varkappa_i'\;,\quad i=1,2,3\;,
\end{equation}
so that $\varkappa_i$ become invariants. Map (\ref{A-map}) is the
Hamiltonian one, it preserves the local symplectic form
\begin{equation}\label{Poisson}
\sum_{i=1}^3 \frac{d\,u_i^{}\wedge d\,w_i^{}}{u_i^{}w_i^{}}\;=\;
\sum_{i=1}^3 \frac{d\,u_i'\wedge d\,w_i'}{u_i'w_i'}\;,
\end{equation}
and with the orthogonality condition (\ref{kappa-inv}) it
satisfies the functional tetrahedron equation
\cite{KashaevKorepanovSergeev}. In what follows, condition
(\ref{orthogonality},\ref{kappa-inv}) is implied.

Thus, due to (\ref{Poisson}), there exists a generation function,
\begin{equation}\label{G}
dG(u;u')\;=\;\sum_{i=1}^3 \left(\log w_i^{\prime}\;d\,\log
u_i^{\prime} \;-\; \log w_i^{}\;d\,\log
u_i^{}\right)_{u_2u_3=u_2'u_3'}\;,
\end{equation}
where $u_i^{\#}$ and $w_i^{\#}$ are related by
(\ref{A-map},\ref{Lambda}). In the definition of generating
function $u_i^{},u_i'$ are chosen as independent variables bounded
by condition $u_2^{}u_3^{}=u_2'u_3'$ following from (\ref{A-map}).

Let $L(z)$ be Roger's dilogarithm,
\begin{equation}
L(z)\;=\;\int_{0}^z \log(1-x) d\log x\;,
\end{equation}
with the branch cut $z\geq 1$. Then the generating function is
given by
\begin{equation}\label{G-answer}
\begin{array}{l}
\ds G(u;u')=
\log\frac{u_3'}{\varkappa_1}\log\frac{u_1'}{u_1}+\log\varkappa_3\log\frac{u_1'}{u_2}
+L(\frac{\varkappa_2}{\varkappa_1}\frac{u_2}{u_1'})
+L(\frac{u_2'}{u_1}) - L(\varkappa_2\frac{u_2'}{u_1'}) -
L(\frac{1}{\varkappa_1}\frac{u_2}{u_1})\\
[1.5mm] \ds = \log
u_3\log\frac{u_1'}{u_1}+\log\varkappa_3\log\frac{u_1'}{u_2} +
\log\varkappa_2\log\frac{u_2}{u_2'} -
L(\frac{\varkappa_1}{\varkappa_2}\frac{u_1'}{u_2}) -
L(\frac{u_1}{u_2'}) + L(\frac{1}{\varkappa_2}\frac{u_1'}{u_2'}) +
L(\varkappa_1\frac{u_1}{u_2})\;.
\end{array}
\end{equation}
Positiveness of $w_i^{\#}$ guarantees that arguments of all
dilogarithms for one of the lines of (\ref{G-answer}) are out of
the branch cut and therefore the generation function is real.

Quantization of local symplectic structure (\ref{Poisson})
$\{u,w\}=uw$ produces the local Weyl algebra
$\uop\wop=q^2\wop\uop$. Quantum counterpart of Hamiltonian form of
(\ref{A-map}) is an intertwiner $R_{123}$ in the tensor cube of
proper representations of local Weyl algebras such that
\begin{equation}\label{intertwiner}
\uop_i'=R_{123}^{}\uop_i^{}R_{123}^{-1}\;,\quad
\wop_i'=R_{123}^{}\wop_i^{}R_{123}^{-1}\;,\quad i=1,2,3\;.
\end{equation}
For instance, the modular representation \cite{Faddeev:1995} of
the local Weyl algebra is given by
\begin{equation}\label{uw-strong}
\uop \;=\;\EXP^{2\pi b \xop}\;,\quad \wop \;=\; - \EXP^{2\pi b
\pop}\;,\quad \varkappa\;=\;-\EXP^{2\pi b\lambda}\;,
\end{equation}
where $\xop,\pop$ is the self-conjugated Heisenberg pair
\begin{equation}\label{Heisenberg}
[\xop,\pop]\;=\;\frac{\ii}{2\pi}\quad\Rightarrow\quad
q\;=\;\EXP^{\ii\pi b^2}\;,
\end{equation}
and ``physical'' regime for $b$ is
\begin{equation}
\eta\;\stackrel{\textrm{def}}{=}\;\frac{b+b^{-1}}{2}\;>\;0\;.
\end{equation}
Modular partner to (\ref{uw-strong}) is
\begin{equation}\label{uw-dual}
\tilde{\uop} \;=\;\EXP^{2\pi b^{-1} \xop}\;,\quad \tilde{\wop}
\;=\; - \EXP^{2\pi b^{-1} \pop}\;,\quad
\tilde{\varkappa}\;=\;-\EXP^{2\pi b^{-1}\lambda}\;.
\end{equation}
Form of the map (\ref{intertwiner}) for
$\tilde{\uop}_i,\tilde{\wop}_i,\tilde{\varkappa}_i$ coincides with
that for $\uop_i,\wop_,\varkappa_i$; in the strong coupling regime
$0<\eta<1$ partner equations are Hermitian conjugated.

Kernel of the intertwiner (\ref{intertwiner}) in the coordinate
representation of Heisenberg pairs (\ref{Heisenberg}) is
\begin{equation}\label{kernel}
\begin{array}{ll}
\ds \langle x_1^{}x_2^{}x_3^{}|R|x_1'x_2'x_3'\rangle \;= & \ds
\delta(x_2^{}+x_3^{}=x_2'+x_3') \EXP^{ 2\pi\ii \left\{
(x_3'-\lambda_1^{})(x_1^{}-x_1') + (\lambda_3^{}-\ii\eta)(x_2^{}-x_1')\right\}}\\
&\\
&\ds \frac{\varphi(x_2^{}-x_1^{}-\lambda_1^{})
\varphi(x_2'-x_1'+\lambda_2^{})}{\varphi(x_2'-x_1^{}-\ii\eta + \ii
\epsilon)\varphi(x_2^{}-x_1'+\lambda_2^{}-\lambda_1^{}-\ii\eta +
\ii \epsilon)}\;,
\end{array}
\end{equation}
where function $\varphi$ is the non-compact quantum dilogarithm
\cite{Faddeev:1995}
\begin{equation}\label{fi-def}
\varphi(z)\;\stackrel{\textrm{def}}{=}\;\exp\left(\ds
\frac{1}{4}\int_{\mathbb{R}+\ii 0} \frac{\EXP^{-2\ii
zw}}{\textrm{sinh}(wb)\textrm{sinh}(w/b)}\ \frac{dw}{w}\right)\;.
\end{equation}
Symbols $\ii\epsilon$ in denominator of (\ref{kernel}) define
circumventions of poles. Operator (\ref{kernel}) satisfies the
quantum tetrahedron equation with free $\lambda_i$.

The choice of negative signs near $\wop$ and $\varkappa$ in
(\ref{uw-strong}) provides the unitarity of operator
(\ref{kernel}) for real $\lambda_i$,
$R_{123}^{-1}=R_{123}^\dagger$. Positive ``geometric'' signs can
be obtained by the analytical continuation
$\lambda_i\to\lambda_i+\ii\eta$ and non-unitary gauge
transformation
$\wop\to\EXP^{-2\pi\eta\xop}\wop\EXP^{2\pi\eta\xop}=-q^{-1}\wop$.
In that case the kernel of $R$-matrix (\ref{kernel}) has the
semi-classical ($b\to 0$ and $\EXP^{2\pi b \xop}\to u$) asymptotic
\begin{equation}
\log\left(\EXP^{-2\pi\eta x_1}\langle x|R|x'\rangle \EXP^{2\pi\eta
x_1'}\right)\; \mathop{\longrightarrow}_{b\to
0}\;-\frac{G(u;u')}{2\pi\ii b^2}\;,
\end{equation}
where the generating function is given by (\ref{G-answer}).

It worth mentioning the cyclic representations of Weyl algebra
with $q^{2N}=1$. The cyclic representation is a $\mathbb{Z}_N$
fiber over the base of centers $\tilde{\uop}=\uop^N$,
$\tilde{\wop}=\wop^N$ \cite{Bazhanov:1995jpa}. Equations of motion
for $\mathbb{C}$-valued centers follow from quantum map
(\ref{A-map}), they just coincide with classical equations of
motion. It is natural then to identify the evoluting centers
$\tilde{\uop}_i,\tilde{\wop}_i$ directly with the geometric data
(\ref{geometry}) and pose quantum problems in Hilbert space
\begin{equation}
\mathcal{H}= \mathbb{Z}_N^{\otimes (\textrm{size of net's
section})}
\end{equation}
in the presence of external classical geometry. The structure of
$\mathbb{Z}_N^{\otimes 3}$ intertwiners and modified tetrahedron
equations are discussed in more details in e.g.
\cite{Tokyo00,GPS03}. Homogeneous point
$\tilde{\uop}_i'=\tilde{\uop}_i^{}$,
$\tilde{\wop}_i'=\tilde{\wop}_i^{}$ of the
Zamolodchikov-Bazhanov-Baxter model \cite{Sergeev:1995rt} is
complex one, it is not a geometrical regime.

\bigskip
\noindent\textbf{Acknowledgements.} I am grateful to V. Bazhanov
and V. Mangazeev for valuable discussions and fruitful
collaboration. Also I would like to thank M. Hewett, P. Vassiliou
and J. Ascione for an encouragement.


\end{document}